\documentclass[10pt,twocolumn]{article}
\usepackage{graphicx}
\usepackage{amsmath,amssymb,times,cite}
\usepackage{booktabs,multirow,xcolor}

\title{\Large\textbf{Synthesis of COVID-19 Chest X-rays using Unpaired Image-to-Image Translation}}
\author{Hasib Zunair and A. Ben Hamza \\
Concordia Institute for Information Systems Engineering \\
Concordia University, Montreal, QC, Canada
}

\date{}

\begin{document}
\maketitle
\begin{abstract}
Motivated by the lack of publicly available datasets of chest radiographs of positive patients with Coronavirus disease 2019 (COVID-19), we build the first-of-its-kind open dataset of synthetic COVID-19 chest X-ray images of high fidelity using an unsupervised domain adaptation approach by leveraging class conditioning and adversarial training. Our contributions are twofold. First, we show considerable performance improvements on COVID-19 detection using various deep learning architectures when employing synthetic images as additional training set. Second, we show how our image synthesis method can serve as a data anonymization tool by achieving comparable detection performance when trained only on synthetic data. In addition, the proposed data generation framework offers a viable solution to the COVID-19 detection in particular, and to medical image classification tasks in general. Our publicly available benchmark dataset\footnote{https://github.com/hasibzunair/synthetic-covid-cxr-dataset} consists of 21,295 synthetic COVID-19 chest X-ray images. The insights gleaned from this dataset can be used for preventive actions in the fight against the COVID-19 pandemic.
\end{abstract}


\section{Introduction}
The World Health Organization (WHO) has declared COVID-19, the infectious respiratory disease caused by the novel coronavirus, a global pandemic due to the rapid increase in infections worldwide. This virus has spread across the globe, sending billions of people into lockdown, as many countries rush to implement strict measures in an effort to slow COVID-19 spread and flatten the epidemiological curve. Although most people with COVID-19 have mild to moderate symptoms, the disease can cause severe lung complications such as viral pneumonia, which is frequently diagnosed using chest radiography.

Recent studies have shown that chest radiography images such as chest X-rays (CXR) or computed tomography (CT) scans performed on patients with COVID-19 when they arrive at the emergency room can help doctors determine who is at higher risk of severe illness and intubation~\cite{ai2020correlation,huang2020clinical}. These X-rays and CT scans show small patchy translucent white patches (called ground-glass opacities) in the lungs of COVID-19 patients. A chest X-ray provides a two-dimensional (2D) image, while a CT scan has the ability to form three-dimensional (3D) images of the chest. However, chest CT based screening is more expensive, not always available at small or rural hospitals, and often yields a high false-positive rate. Therefore, the need to develop computational approaches for detecting COVID-19 via chest radiography imaging not only can save healthcare a tremendous amount of time and money, but more importantly, it can save more lives~\cite{ng2020imaging}. By leveraging deep learning, several approaches for the detection of COVID-19 cases from chest radiography images have been recently proposed, including tailored convolutional neural network (CNN) architectures~\cite{karim2020deepcovidexplainer,wang2020covid} and transfer learning based methods~\cite{kassani2020automatic,narin2020automatic,XinLi:20,Farooq:20}.

While promising, the predictive performance of these deep learning based approaches depends heavily on the availability of large amounts of data. However, there is a significant shortage of chest radiology imaging data for COVID-19 positive patients, due largely to several factors, including the rare nature of the radiological finding, legal, privacy, technical, and data-ownership challenges. Moreover, most of the data are not accessible to the global research community.

In recent years, there have been several efforts to build large-scale annotated datasets for chest X-rays and make them publicly available to the global research community~\cite{demner2016preparing,johnson2019mimic,irvin2019chexpert,wang2017chestx,bustos2019padchest}. At the time of writing, there exists, however, only one annotated COVID-19 X-ray image dataset~\cite{cohen2020covid}, which is a curated collection of X-ray images of patients who are positive or suspected of COVID-19 or other viral and bacterial pneumonias. This COVID-19 image data collection has been used as a primary source for positive cases of COVID-19~\cite{karim2020deepcovidexplainer,wang2020covid,kassani2020automatic,narin2020automatic}, where the detection of COVID-19 is formulated as a classification problem. While the COVID-19 image data collection contains positive examples of COVID-19, the negative examples were acquired from publicly available sources~\cite{wang2017chestx} and merged together for data-driven analytics. This fusion of multiple datasets results in predominantly negative examples with only a small percentage of positive ones, giving rise to a class imbalance problem~\cite{demner2016preparing,johnson2019mimic,irvin2019chexpert,wang2017chestx,bustos2019padchest}. This in turn becomes a challenge of its own, as the merged data becomes highly imbalanced. In the context of a classifier training, the class imbalance problem in the training data distribution yields sub-optimal performance on the minority class (i.e. positive class for COVID-19).

\begin{figure}[!htb]
\centering
\includegraphics[width=3.4in, height=1.7in]{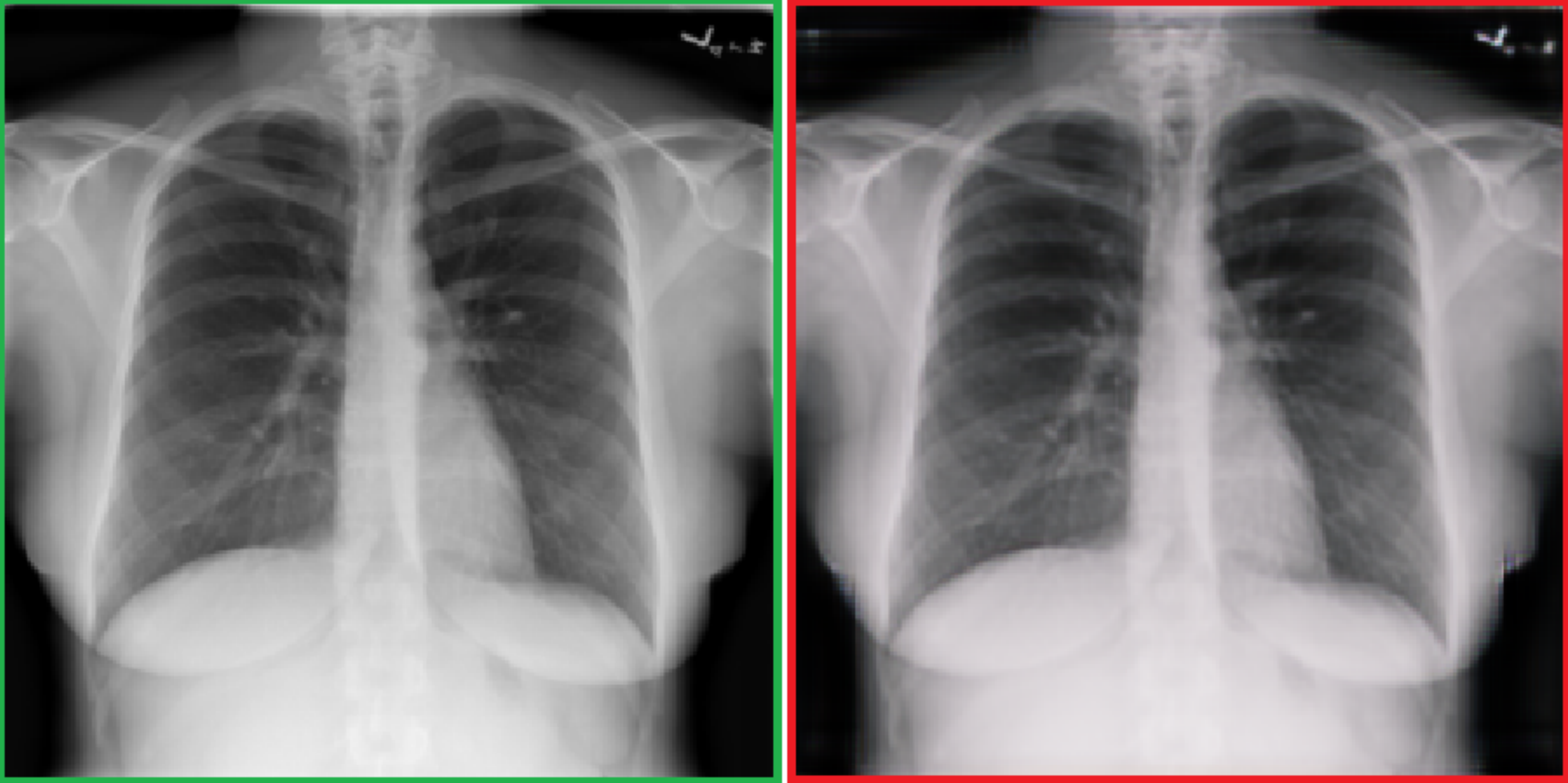}\\[.6ex]
\includegraphics[width=3.4in, height=1.7in]{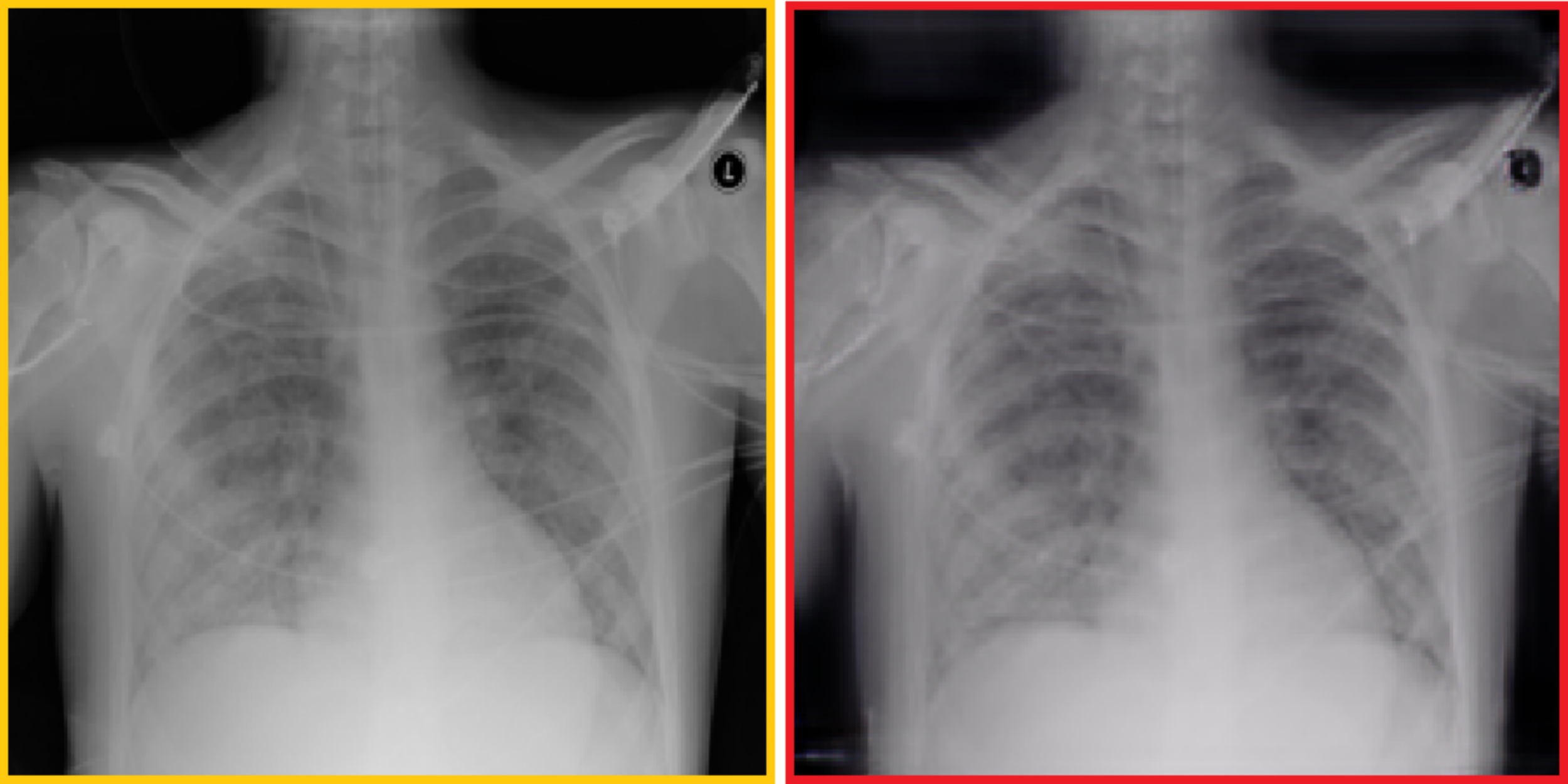}
\caption{Given any two unordered image collections, the generative algorithm learns to automatically translate an image from one category to another. Top: Normal (left) to COVID-19 (right). Bottom: Pneumonia (left) to COVID-19 (right). Subtle visual changes can be observed in the translated images due to low inter-class variation.}
\label{Fig:translation_demo}
\end{figure}

In order to overcome the aforementioned issues, we present a domain adaptation framework by leveraging the inter-class variation of the data distribution for the task of conditional image synthesis by learning the inter-class mapping and synthesizing under-represented class samples from the over-represented ones using unpaired image-to-image translation~\cite{zhu2017unpaired}. The proposed framework combines class-conditioning and adversarial training in a bid to synthesize realistic looking COVID-19 CXR images. The generated synthetic dataset contains 21,295 synthetic images of chest X-rays for COVID-19 positive cases. Figure~\ref{Fig:translation_demo} shows how our approach learns to automatically translate an image from one category to another, and more specifically from normal to COVID-19, and pneumonia to COVID-19.

In addition to demonstrating improved COVID-19 detection performance through the use of various deep convolutional neural network architectures on the synthetic data to boost training, we show how the proposed data generation and evaluation pipeline can serve as a viable data-driven solution to medical image analysis problems, and make our dataset publicly available, which is currently comprised of 21,295 synthetic images of chest X-rays for COVID-19 positive cases. The main contributions of this paper can be summarized as follows:
\begin{itemize}
\item We present an integrated deep learning based framework, which couples adversarial training and transfer learning to jointly address inter-class variation and class imbalance.
\item We synthesize chest X-ray images of COVID-19 to adjust the skew in training sets by over-sampling positive cases to mitigate the class imbalance problem, while training classifiers.
\item We demonstrate how the data generation procedure can serve as an anonymization tool by achieving comparable detection performance when trained only on synthetic data versus real data in an effort to alleviate privacy concerns.
\end{itemize}

\medskip\noindent
The rest of this paper is organized as follows. In Section 2, we provide a brief overview of generative approaches for medical image synthesis. In Section 3, we present a generative framework, which couples adversarial training and transfer learning to jointly address inter-class variation and class imbalance. In Section 4, we present experimental results to demonstrate improved COVID-19 detection performance through the use of various deep convolutional neural network architectures on the generated data. Finally, we conclude in Section 5 and point out future work directions.

\section{Related Work}
The advent of generative adversarial networks (GANs)~\cite{goodfellow2014generative} has accelerated research in generative modeling and distribution learning. With the ability to replicate data distributions and synthesize images with high fidelity, GANs have bridged the gap between supervised learning and image generation. These synthetic images can then be used as input to improve the performance of various deep learning algorithms for downstream tasks, such as image classification and segmentation. GANs have not only been used in natural images' settings, but have also been extensively employed in medical image analysis~\cite{kazeminia2018gans}, where labels are usually scarce or almost non-existent.

With the scarcity of annotated medical image datasets, there has been a surge of interest in developing efficient approaches for the generation of synthetic medical images. While several existing generative methods have addressed the translation between multiple imaging modalities CT-PET, CS-MRI, MR-CT, XCAT-CT~\cite{ben2017virtual,yang2017dagan,wolterink2017deep,russ2019synthesis} based on distribution matching, other approaches have focused on the scarcity of labeled data in the medical field due in large part to the acquisition, privacy and health safety issues. Conditional and unconditional image synthesis procedures, built on top of these generative models, have been proposed in retinal images~\cite{costa2017towards,dar2019image} and MRI scans~\cite{shin2018medical,guibas2017synthetic,korkinof2018high}. These models involve the training of paired data in both source and target domains to synthesize realistic, high-resolution images in order to aid in medical image classification and segmentation tasks.

Image synthesis methodologies have also been proposed in the context of chest X-rays~\cite{teixeira2018generating}. Our work is significantly different in the sense that we are specifically interested in synthesizing a particular class, whereas in~\cite{teixeira2018generating} X-rays are generated from surface geometry for landmark detection tasks. While some generative methods only require paired data in the source domain with target domain consisting of unlabelled examples, Cohen \textit{et al.}~\cite{cohen2018distribution} have demonstrated that the phenomenon of \emph{hallucinating features} (e.g. adding or removing tumors leading to a semantic change) leads to a high bias in these domain adaptation techniques. To overcome this issue, we have recently proposed a domain adaptation technique based on cycle-consistent adversarial networks in order synthesize high fidelity positive examples to improve detection performance of melanoma from skin lesion images~\cite{zunair2020melanoma}.

\section{Proposed Method} \label{sec:method}
In this section, we present the main building blocks of our proposed image synthesis framework using image-to-image translation, which is an increasingly popular machine learning paradigm that has shown great promise in a wide range of applications, including computer graphics, style transfer, satellite imagery, object transfiguration, character animation, and photo enhancement. In an typical image-to-image translation problem, the objective is to learn a mapping that translates an image in one domain to a corresponding image in another domain using approaches that leverage paired or unpaired training samples. The latter is the focus of our work. While paired image-to-image translation methods use pairs of corresponding images in different domains, the paired training samples are, however, not always available. By contrast, the unpaired image-to-image translation problem, in which training samples are readily available, is more common and practical, but it is highly under-constrained and fraught with challenges. In our work, we build upon the idea that there exist no paired training samples showing how an image from one domain can be translated to a corresponding image in another domain. The task is to generate COVID-19 chest X-rays from chest X-ray images to address COVID-19 class imbalance problem. More specifically, our goal is to learn a mapping function between Non-COVID-19 images and COVID-19 in order to generate COVID-19 chest X-rays without paired training samples in an unsupervised fashion.

\subsection{Chest X-ray image synthesis}
We formulate the detection of COVID-19 as a binary classification problem. For the Normal vs. COVID-19 and Pneumonia vs. COVID-19 tasks, we train two translation models and synthesize COVID-19 images for each task in order to adjust the skew in the training data by over-sampling the minority class. For the sake of clarity and unless otherwise expressly indicated, we refer to the source domain of the two tasks as \emph{Non-COVID-19} instead of \emph{Normal} and \emph{Pneumonia} separately.

We adopt our unsupervised domain adaptation technique introduced in~\cite{zunair2020melanoma} to translate Non-COVID-19 images for each case (i.e. normal or pneumonia) to COVID-19. Given two image domains $A$ and $B$ denoting Non-COVID-19 and COVID-19, respectively, the goal is to learn to translate images of one type to another using two generators $G_{A}: A\rightarrow B$ and $G_{B}: B\rightarrow A$, and two discriminators $D_{B}$ and $D_{A}$, as illustrated in Figure \ref{Fig:architecture}.

The generator $G_{A}$ (resp. $G_{B}$) translates images from Non-COVID-19 to COVID-19 (i.e. $A\rightarrow B$), while the discriminator $D_{B}$ (resp. $D_{A}$) verifies how real an image of $B$ (resp. $A$) looks. The overall objective function is defined as
\begin{equation}
\begin{split}
\mathcal{L}(G_{A}, G_{B}, D_{B}, D_{A}) &= \mathcal{L}_{GAN}(G_{A}, D_{B}, A, B) \\
&\quad+ \mathcal{L}_{GAN}(G_{B}, D_{A}, B, A) \\
&\quad+ \lambda\,\mathcal{L}_{cyc}(G_{A}, G_{B}),
\end{split}
\label{eq:cyclegan_loss}
\end{equation}
which consists of two adversarial losses and a cycle consistent loss regularized by a hyper-parameter $\lambda$\cite{zhu2017unpaired}. The first adversarial loss is given by
\begin{equation}
\begin{split}
\mathcal{L}_{GAN}(G_{A},D_{B},A,B) &= \mathbb{E}_{b\sim p_{\text{data}}(b)}[\log D_{B}(b)] \\
&\hspace*{-.4in} + \mathbb{E}_{a\sim p_{\text{data}}(a)}[\log( 1 - D_{B}(G_{a}(a)))],
\end{split}
\label{eq:advloss}
\end{equation}
where the generator $G_A$ tries to generate images $G_{A}(a)$ that look similar to COVID-19 images, while $D_B$ aims to distinguish between generated samples $G_{A}(a)$ and real samples $b$. During the training, as $G_{A}$ generates a COVID-19 image, $D_{B}$ verifies if the translated image is actually a real COVID-19 image or a synthetic one. The data distributions of Non-COVID-19 and COVID-19 are $p_{\text{data}}(a)$ and $p_{\text{data}}(b)$, respectively.

Similarly, the second adversarial loss is given by
\begin{equation}
\begin{split}
\mathcal{L}_{GAN}(G_{B},D_{A},B,A) &= \mathbb{E}_{a\sim p_{\text{data}}(a)}[\log D_{A}(a)] \\
&\hspace*{-.4in} + \mathbb{E}_{b \sim p_{\text{data}}(b)}[\log( 1 - D_{A}(G_{B}(b)))],
\end{split}
\label{eq:advloss2}
\end{equation}
where $G_B$ takes a COVID-19 image $b$ from $B$ as input, and tries to generate a realistic image $G_{B}(b)$ in $A$ that tricks the discriminator $D_B$. Hence, the goal of $G_{B}$ is to generate a Non-COVID-19 chest X-ray such that it fools the discriminator $D_{A}$ to label it as a real Non-COVID-19 image.

\noindent The third loss term is to enforce cycle consistency an is given by
\begin{equation}
\begin{split}
\mathcal{L}_{cyc} (G_{A}, G_{B}) &= \mathbb{E}_{a \sim p_{\text{data}} (a)} [\|G_{B}(G_{A}(a)) - a\|_1] \\
&\,\,+ \mathbb{E}_{b \sim p_{\text{data}} (b)} [\|G_{A}(G_{b}(b)) - b\|_1],
\end{split}
\label{eq:cycleloss}
\end{equation}
which computes the difference between the input image and the generated one using the $\ell_1$-norm.

\subsection{Model optimization}
The idea of the cycle consistency loss it to add a constraint such that $G_{B}(G_{A}(a))\approx a$ and $G_{A}(G_{B}(b))\approx b$. In other words, the objective is to learn two bijective generator mappings by solving the following optimization problem:
\begin{equation}
G_{A}^{\ast}, G_{B}^{\ast}=\arg\min_{G_{A},G_{B}}\max_{D_{A},D_{B}}\mathcal{L}(G_{B}, G_{B}, D_{B}, D_{A}).
\end{equation}
For the generators $G_{A}$ and $G_{B}$, the architecture is based on fully convolutional network (FCN). The discriminators $D_{B}$ and $D_{A}$ consists of a CNN classifier which verifies whether the image is real or synthetic.

\subsection{Training procedure}
The training for the generators and discriminators are carried out in the same way as in~\cite{zunair2020melanoma}. First, we balance the inter-class data samples by performing undersampling. Then, we train Cycle-GAN to learn a function of the interclass variation between the two groups, i.e. we learn a transformation between Non-COVID-19 and COVID-19 radiographs. We apply CycleGAN to the over-represented class samples in order to synthesize the target class samples (i.e. under-represented class).

After training, we apply the generators $G_{A}$ and $G_{B}$ on the training datasets of Normal vs. COVID-19 and Pneumonia vs. COVID-19. We apply $G_{A}$ on the majority class of Normal vs. COVID-19, which consists of normal images in order to synthesize 16,537 COVID-19 images. We denote this synthesized dataset as $\mathcal{G}_{NC}$, which consists of generated images by performing image-to-image translation from normal to COVID-19.

\begin{figure}[!htb]
\centering
\includegraphics[scale=.35]{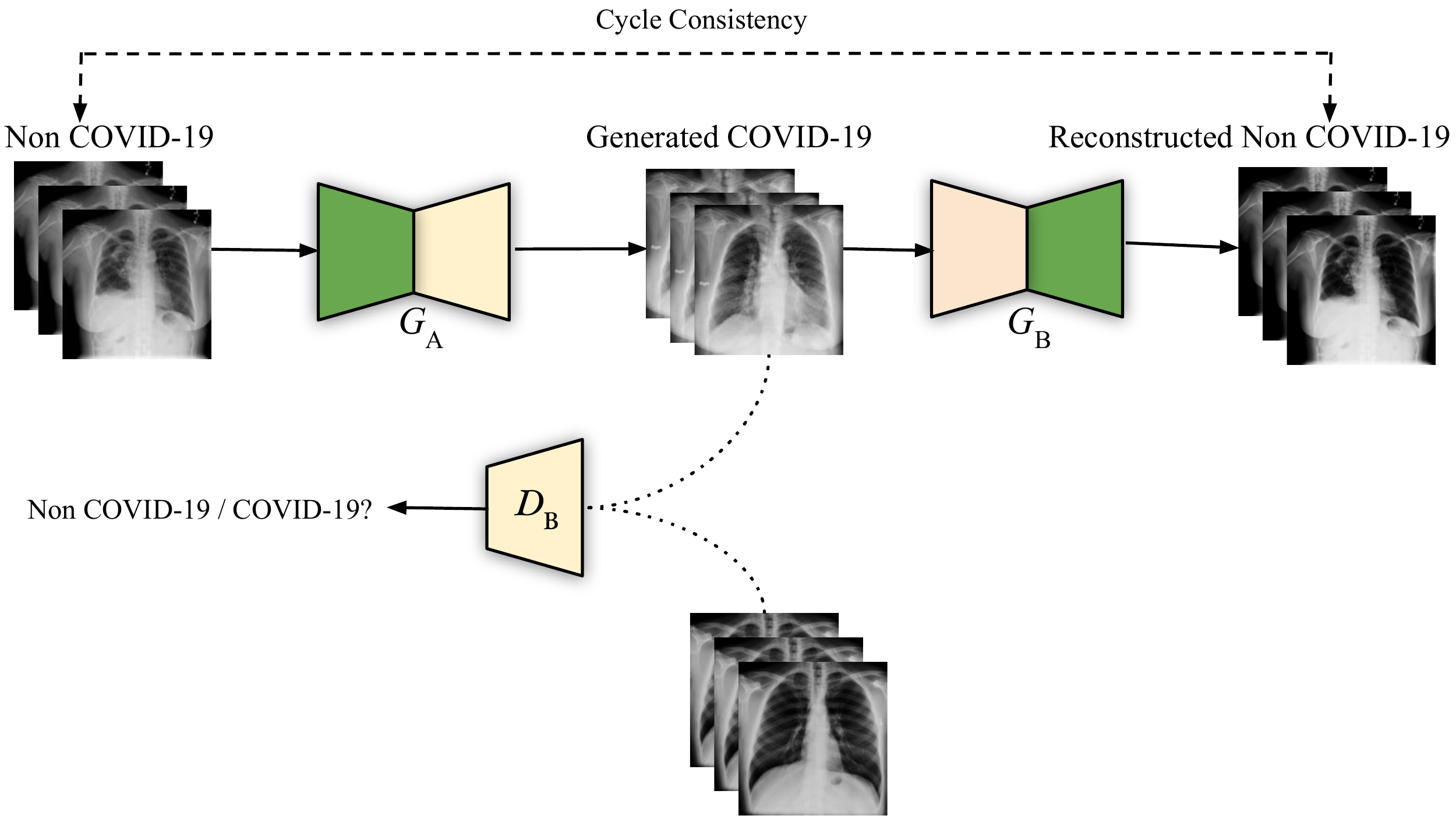}
\caption{Illustration of the generative adversarial training process for unpaired image-to-image translation. Chest X-ray images are translated from Non-COVID-19 (i.e. Normal or Pneumonia) to COVID-19 and then back to Non-COVID-19 to ensure cycle consistency in the forward pass. The same procedure is applied in the backward pass from COVID-19 to Non-COVID-19.}
\label{Fig:architecture}
\end{figure}

Similarly, for Pneumonia vs. COVID-19, we synthesize 4,758 COVID-19 images by applying $G_{B}$ on the majority class consisting of pneumonia images and we denote the synthesized dataset as $\mathcal{G}_{PC}$, which is comprised of generated images by performing image-to-image translation from pneumonia to COVID-19. It is worth pointing out that for the sake of clarity, the discriminator $D_A$ is not depicted to Figure~\ref{Fig:architecture}, as our main is to generate COVID-19 images from Non COVID-19 images.

\section{Experiments} \label{Method}
In this section, we conduct extensive experiments to evaluate the performance of the proposed data generation framework on COVID-19 detection.
\subsection{Datasets}
We use two publicly available datasets of chest X-rays:

\medskip\noindent\textbf{COVID-19 Image Data Collection.}\quad This dataset comprises 226 images of pneumonia cases with chest X-ray or CT images, specifically COVID-19 cases as well as MERS, SARS, and ARDS. Data are scraped from publications and websites such as Radiopaedia.org, Italian Society of Medical and Interventional Radiology\footnote{https://www.sirm.org/category/senza-categoria/covid-19/\label{rd}}, and Figure1.com\footnote{https://www.figure1.com/covid-19-clinical-cases\label{fig}}. From this dataset, we discard the CT images and retain the 226 images positive for COVID-19 and their corresponding labels.

\medskip\noindent\textbf{RSNA Pneumonia Detection Challenge.}\quad This dataset originated from a Kaggle challenge\footnote{https://www.kaggle.com/c/rsna-pneumonia-detection-challenge\label{kaggle}} and consists of publicly available data from~\cite{wang2017chestx}. It is composed of 26684 images, and each image was annotated by a radiologist for the presence of lung opacity; thereby providing a label for two classes. This label is included as both lung opacity and pneumonia.

\subsection{Dataset splits and preprocessing}  \label{sec:datasplit}
We partition the three classes from COVID-19 Image Data Collection and RSNA Pneumonia Detection Challenge into two sets, namely ``Normal vs. COVID-19'' and ``Pneumonia vs COVID-19''. A patient level split is then applied using 80\% as training set and the remaining 20\% as test set to assess algorithm performance, and we follow the same evaluation protocol laid out in~\cite{shin2018medical,zunair2020melanoma}. We define the skew ratio as follows:
\begin{equation}
\text{Skew} = \frac{\text{Negative Examples}}{\text{Positive Examples}},
\label{eq:skew}
\end{equation}
where $\text{Skew} = 1$ represents a fully balanced dataset, $\text{Skew} > 1$ shows that the negative samples are the majority, and $\text{Skew} < 1$ represents positive sample dominance in the distribution.

\begin{figure}[!htb]
\centering
\includegraphics[width=\linewidth]{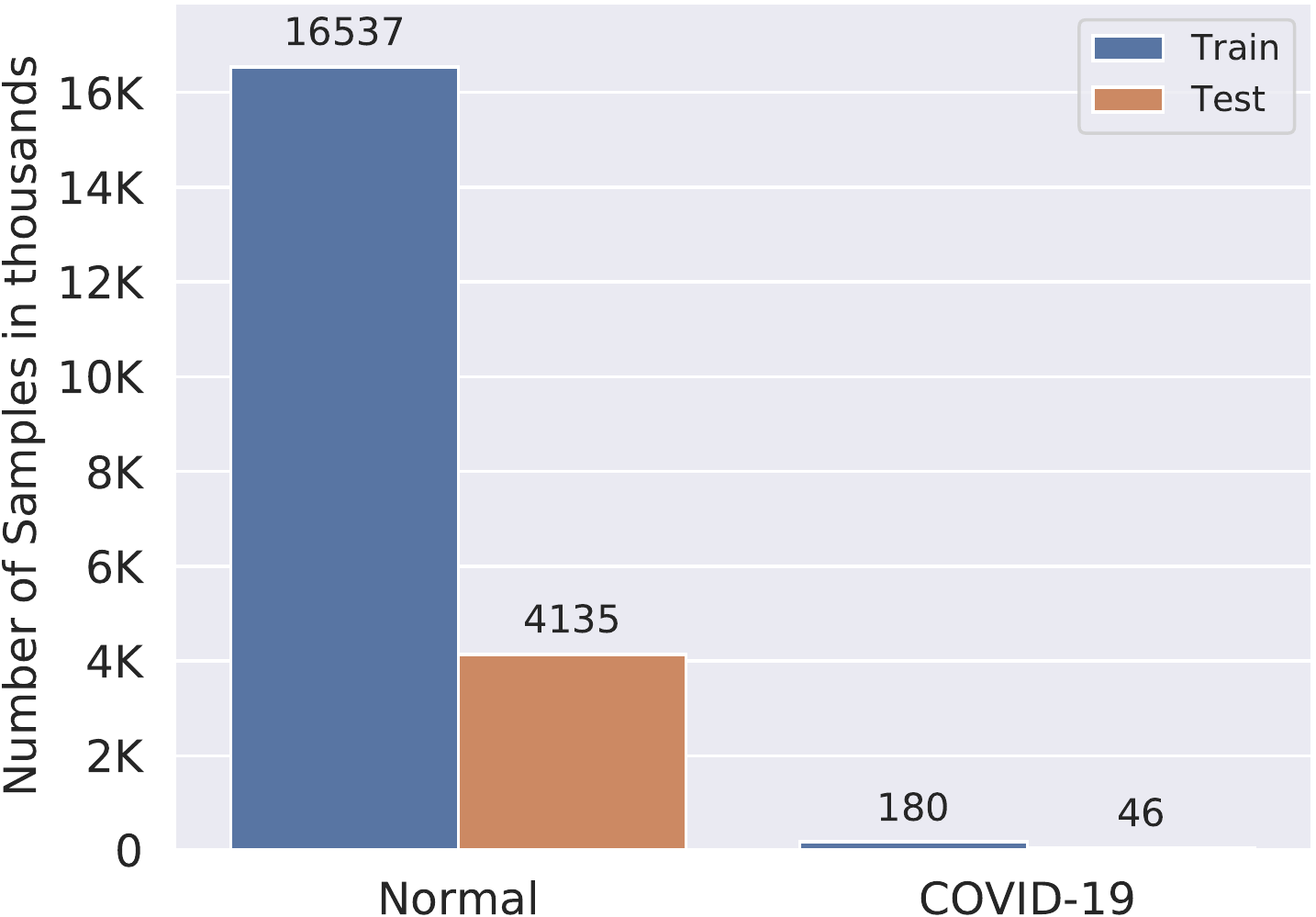}\\[2ex]
\includegraphics[width=\linewidth]{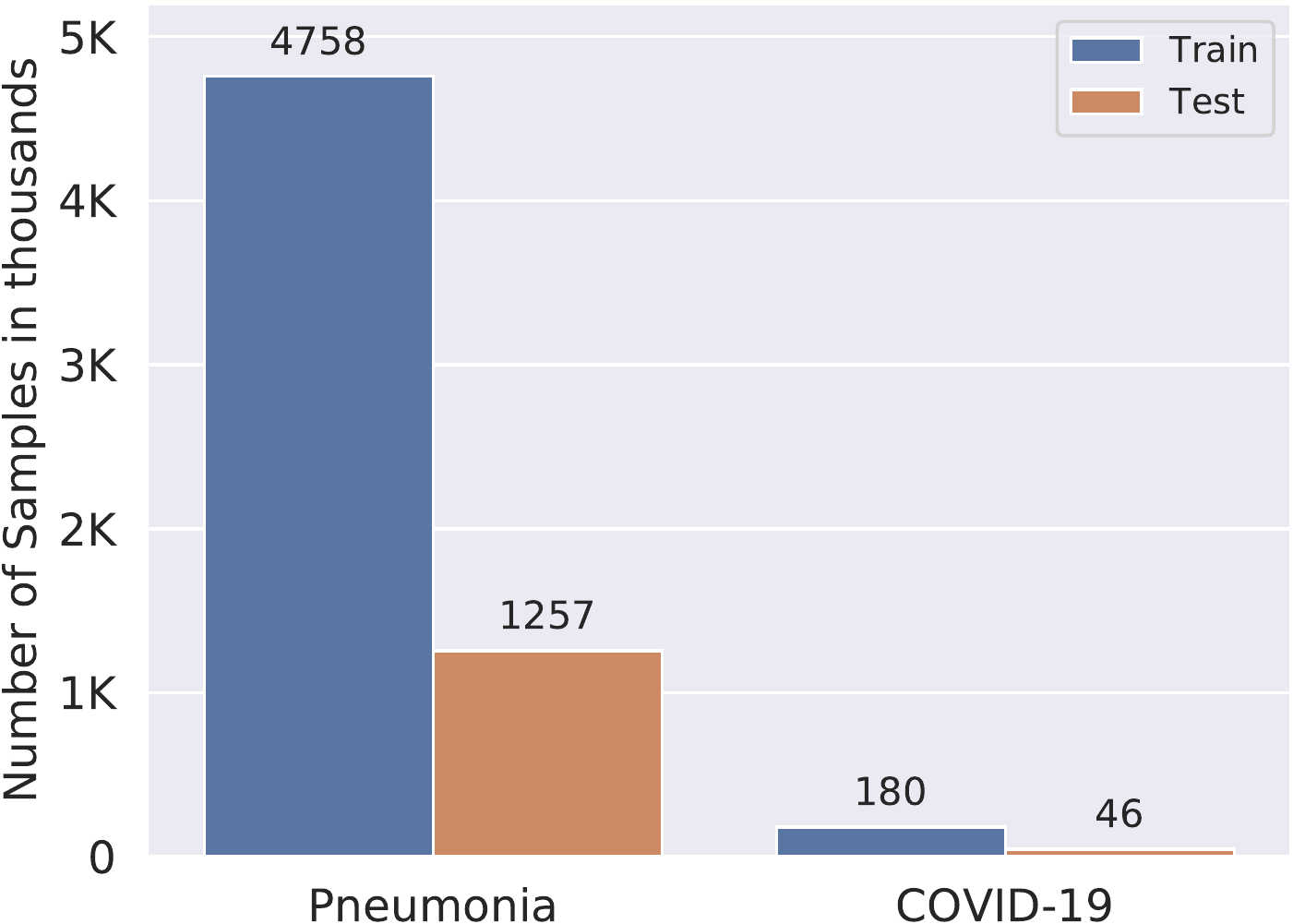}
\caption{Data distributions of Normal vs. COVID-19 (top) and Pneumonia vs. COVID-19 (bottom) with skew ratios of 91.87 and 22.9, respectively.}
\label{Fig:histograms}
\end{figure}

The data distributions of Normal vs. COVID-19 and Pneumonia vs. COVID-19 are displayed in Figure~\ref{Fig:histograms}, which illustrates the class imbalance in the training dataset. For Pneumonia vs. COVID-19, the skew ratio is around 22.9, while the skew for Normal vs. COVID-19 is almost four times larger, indicating high imbalance in the classes.

We also resize all images to $256\times256$ pixels, and scale the pixel values to $[0,1]$ for the training of classifiers. It is important to mention that when we use the term \emph{synthetic data}, we refer to COVID-19 CXR images only.

\subsection{Baselines}
Since our aim is to provide a dataset to be used as a training set for the minority class, we test the effectiveness of several deep CNN architectures, including VGG-16\cite{simonyan2014very}, ResNet-50\cite{he2016deep} and DenseNet-102~\cite{huang2017densely}, on the detection of the minority class. These pretrained networks were trained on more than a million images from the ImageNet database\footnote{http://www.image-net.org}. More specifically, we investigate the contribution of the synthetic datasets $\mathcal{G}_{NC}$ and $\mathcal{G}_{PC}$, which consist of COVID-19 CXR images, to the overall performance of these deep learning models. The last layer of each of these models consists of a global average pooling (GAP) layer, which computes the average output of each feature map in the previous layer and helps minimize overfitting by reducing the total number of parameters in the model. The GAP layer turns a feature map into a single number by taking the average of the numbers in that feature map. Similar to max-pooling layers, GAP layers have no trainable parameters and are used to reduce the spatial dimensions of a 3D tensor. The GAP layer is followed by a single fully connected (FC) layer with a softmax function (i.e. a dense softmax layer of two units for the binary classification case), which yields the predicted classes' probabilities that sum to one.
\subsection{Evaluation metrics}
Due to high class imbalance in the datasets, the choice of evaluation metrics plays a vital role in the comparison of classifiers. Threshold metrics such as accuracy and rank metrics (e.g. area under the ROC curve) may lead to a false sense of superiority and mask poor performance~\cite{jeni2013facing}, thereby introducing bias. Since we are interested in the detection of the minority class (COVID-19), we follow the recommendations provided in~\cite{brabec2018bad,jeni2013facing} and perform quantitative evaluations using sensitivity and false negatives in the same vein as~\cite{kassani2020automatic}. Sensitivity is the percentage of positive instances correctly classified and is defined as
\begin{equation}
\text{Sensitivity} = \frac{\text{TP}}{\text{TP}+\text{FN}},
\label{eq:sens}
\end{equation}
where TP, FP, TN and FN denote true positives, false positives, true negatives and false negatives, respectively. TP is the number of correctly predicted malignant lesions, while TN is the number of correctly predicted benign lesions. A classifier that reduces FN (ruling COVID-19 out in cases that do have it) and FP (wrongly diagnosing COVID-19 where there is none) indicates a better performance. A false negative COVID-19 result can be a serious problem due to the fact that we lose the benefits of early intervention. A false positive result can also cause significant issues for both an individual and the community. Even from an epidemiologicial perspective, a high number of false positives can lead to a wrong understanding of the spread of COVID-19 in the community. Sensitivity, also known as recall or true positive rate, indicates how often a classifier misses a positive prediction. It is one of the most common measures to evaluate a classifier in  medical image classification tasks~\cite{esteva2017dermatologist}. A larger value of Sensitivity indicates a better performance of the classification model.

\subsection{Implementation details}
All experiments are performed on a Linux Workstation (CPU: AMD 2nd Gen Ryzen Threadripper 2950X, 16-Core, 64-Thread, 4.4GHz Max Boost; Memory: 64GB high-performance RAM; GPU: NVIDIA GeForce RTX 2080 Ti). We perform training/testing on both COVID-19 Image Data Collection and RSNA Pneumonia Detection Challenge. For training the models, we use the Adadelta optimization alogrithm~\cite{zeiler2012adadelta} to minimize the binary cross-entropy loss function with a learning rate of $0.001$ and batch size of 16. We initialize the weights using ImageNet and train all layers until the loss stagnates using an early stopping mechanism.

For each dataset, we follow the same evaluation protocol laid out in~\cite{shin2018medical,zunair2020melanoma} for testing the contribution of newly added data. In this evaluation protocol, both training and test sets are used. The training set varies, as new data are added to each configuration. The deep CNN classifiers are trained on this data and evaluated on the held-out test set. For fair evaluation and comparison purposes, the size of the test set remains constant. It is important to mention that the test set does not contain any synthetic examples. Moreover, the hyper-parameters are not tuned and hence do not require a separate validation set.
\subsection{Over-sampling with synthetic data}
We demonstrate the effectiveness of the synthetic sets $\mathcal{G}_{NC}$ and $\mathcal{G}_{PC}$ in Tables~\ref{Tab:class} and~\ref{Tab:synthetic} using four deep learning models, namely VGG-16~\cite{simonyan2014very}, ResNet-50~\cite{he2016deep}, DenseNet-102~\cite{huang2017densely}, and DenseNet-121 with a bagging tree classifier (DenseNet121 + BGT)~\cite{kassani2020automatic}. For each task, we can observe that when $\mathcal{G}_{NC}$ is added, there is a significant increase in performance. While the addition of $\mathcal{G}_{PC}$ also results in an increase in performance, such an increase is not quite large compared to adding $\mathcal{G}_{NC}$ in some cases. We hypothesize that this is due to the number of COVID-19 examples in $\mathcal{G}_{NC}$ (16,537), which enables the models to learn better representations for COVID-19, whereas $\mathcal{G}_{PC}$ is comprised of only 4,758 COVID-19 examples. Further, an increase in performance using both metrics is observed when the skew in the training dataset decreases. The relative improvement seems to drop as the model complexity increases, which is in line with the findings in~\cite{raghu2019transfusion} due to the problem of over-parametrization. When synthetic data are used as additional training set, the detection performance significantly increases. However, the relative improvement drops when the architectural complexity of the model increases. Note that despite its simplicity, the VGG-16 network outperforms all the other baseline methods, while the ``DenseNet121 + BGT'' model yields the second best performance. For less complex models, we can see that using only synthetic dataset performs better than the original data. Moreover, Table~\ref{Tab:synthetic} shows that with the exception of VGG-16, all models achieve sub-optimal performance when using synthetic data only.

\begin{table*}[!htb]
\caption{COVID-19 detection performance results on Normal vs. COVID-19 test set when trained on real data; real + synthetic data; and only synthetic data (i.e. only $\mathcal{G}_{NC}$ is used for positive class examples in training each model). SEN is short for Sensitivity. Boldface numbers indicate the best performance.}
\label{Tab:class}
\setlength{\tabcolsep}{4.2pt}
\centering
\medskip
\begin{tabular}{@{}lccccccccccc@{}}
 \toprule
 & \multicolumn{3}{c}{Real} & \multicolumn{3}{c}{Real + $\mathcal{G}_{NC}$} & \multicolumn{3}{c}{Real + $\mathcal{G}_{NC}$ + $\mathcal{G}_{PC}$} & \multicolumn{2}{c}{Only Synthetic} \\
 \cmidrule(lr){2-4} \cmidrule(lr){5-7} \cmidrule(lr){8-10} \cmidrule(lr){11-12}
 Model & SEN (\%) $\uparrow$ & FN $\downarrow$ & Skew$\downarrow$ & SEN (\%) $\uparrow$ & FN $\downarrow$ & Skew$\downarrow$ & SEN (\%) $\uparrow$ & FN $\downarrow$ & Skew$\downarrow$ & SEN (\%) $\uparrow$ & FN $\downarrow$ \\
 \midrule
 VGG-16  & 19.56 & 37 & 91.87 & 54.34 & 21 & 0.98 & \textbf{63.04} & \textbf{17} & 0.79 & 50.00 & 23 \\
 ResNet-50  & 32.61 & 31 & 91.87 & 41.30 & 27 & 0.98 & \textbf{43.47} & \textbf{26} & 0.79 & 10.86 & 41 \\
 DenseNet-102  & 26.08 & 34 & 91.87 & 28.27 & 33 & 0.98 & \textbf{34.73} & \textbf{30} & 0.79 & 8.69 & 42 \\
 DenseNet-121 + BGT  & 36.95 & 29 & 91.87 & 45.65 & 25 & 0.98 & \textbf{52.17} & \textbf{22} & 0.79 & 21.73 & 36 \\ \bottomrule
\end{tabular}
\end{table*}

\begin{table*}[!htb]
\caption{COVID-19 detection performance results on Pneumonia vs. COVID-19 test set when trained on real data; real + synthetic data; and only synthetic data (i.e. only $\mathcal{G}_{PC}$ is used for positive class examples in training each model). Boldface numbers indicate the best performance.}
\label{Tab:synthetic}
\setlength{\tabcolsep}{4.2pt}
\centering
\medskip
\begin{tabular}{@{}lccccccccccc@{}}
 \toprule
  & \multicolumn{3}{c}{Real} & \multicolumn{3}{c}{Real + $\mathcal{G}_{PC}$} & \multicolumn{3}{c}{Real + $\mathcal{G}_{PC}$ + $\mathcal{G}_{NC}$} & \multicolumn{2}{c}{Only Synthetic} \\
 \cmidrule(lr){2-4} \cmidrule(lr){5-7} \cmidrule(lr){8-10} \cmidrule(lr){11-12}
 Model & SEN (\%) $\uparrow$ & FN $\downarrow$ & Skew$\downarrow$ & SEN (\%) $\uparrow$ & FN $\downarrow$ & Skew$\downarrow$ & SEN(\%) $\uparrow$ & FN $\downarrow$ & Skew$\downarrow$ & SEN (\%) $\uparrow$ & FN $\downarrow$ \\
 \midrule
 VGG-16 & 8.69 & 42 & 22.9 & 29.50 & 24 & 0.95 & \textbf{52.17} & \textbf{22} & 0.19 & 39.13 & 28 \\
 ResNet-50 & 21.73 & 36 & 22.9 & 36.95 & 29 & 0.95 & \textbf{41.30} & \textbf{27} & 0.19 & 13.04 & 40 \\
 DenseNet-102 & 4.34 & 44 & 22.9 & 21.74 & 36 & 0.95 & \textbf{32.43} & \textbf{32} & 0.19 & 6.52 & 43 \\
DenseNet-121 + BGT & 32.60 & 31 & 22.9 & 41.30 & 27 & 0.95 & \textbf{47.82} & \textbf{24} & 0.19 & 32.60 & 31 \\ \bottomrule
\end{tabular}
\end{table*}

\subsection{Training on anonymized synthetic data} \label{privacy}
We also evaluate the performance when the classifiers are trained on only synthetic COVID-19 images, as shown in Tables~\ref{Tab:class} and~\ref{Tab:synthetic} for each dataset. Sub-optimial performance is achieved for both tasks for different CNNs, except for VGG-16 which shows performance improvement compared to when using the original COVID-19 examples. Since a new data sample is not attributed to an individual patient, but it is rather an instance which is conditioned on the training data, it does not entirely reflect the original data. This suggests that synthetic data alone cannot to used to achieve optimal performance. In other words, the synthetic data can be used as a form of pre-training, which often requires a small amount of real data to achieve comparable performance. In addition, the relatively large margin between the evaluation scores suggests that the observed difference between the models is actually real, and not due to a statistical chance.

\subsection{Detecting target class with high confidence}
\begin{figure}[!htb]
\setlength{\tabcolsep}{.1em}
\centering
\begin{tabular}{cc}
\includegraphics[scale=.31]{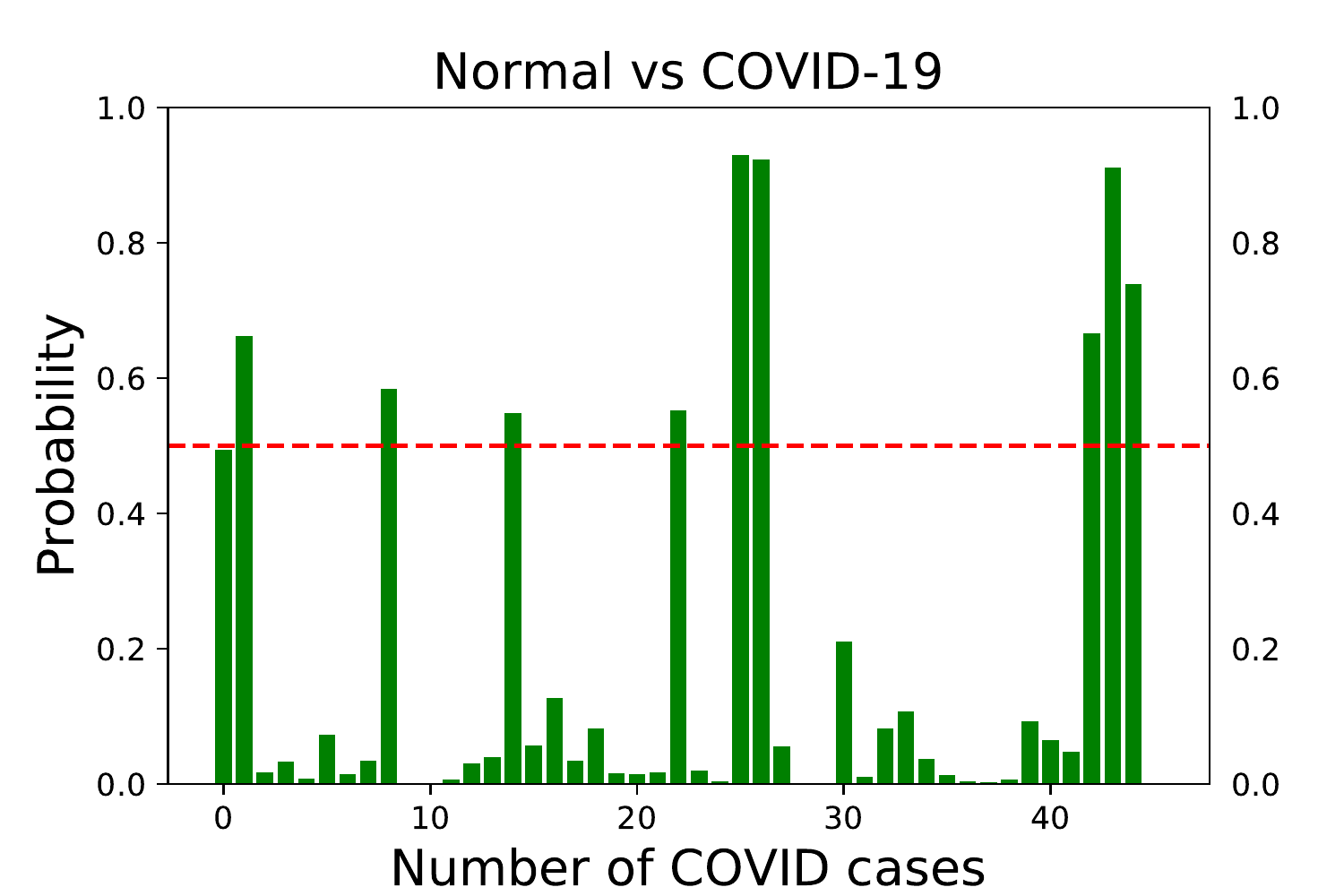} & \includegraphics[scale=.31]{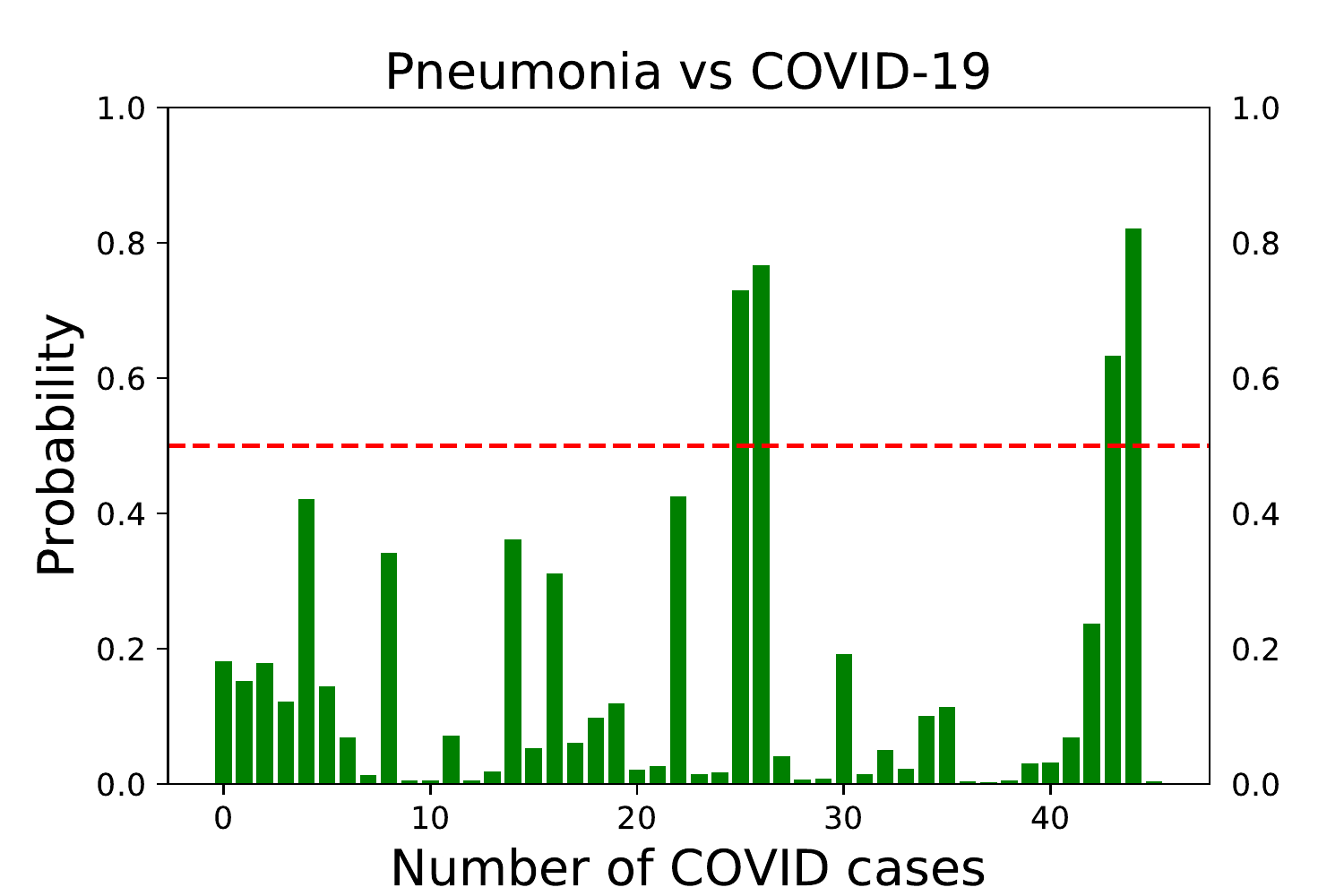}
\end{tabular}
\caption{Confidence scores for the VGG-16 model on unseen test set of COVID-19 for the two binary classification tasks of Normal vs. COVID-19 (left) and Pneumonia vs. COVID-19 (right) using original data only.}
\label{fig:probplots1}
\end{figure}

The output of the softmax function describes the probability (or confidence) of the learning model that a particular sample belongs to a certain class. The softmax layer takes the raw values (logits) of the last FC layer and maps them into probability scores by taking the exponents of each output and then normalize each number by the sum of those exponents so that all probabilities sum to one. Figure~\ref{fig:probplots1} shows the probability scores for the VGG-16 model on unseen test set of COVID-19 for the two binary classification tasks of Normal vs. COVID-19 and Pneumonia vs. COVID-19 using original data only. The red dashed line depicts the 0.5 probability threshold. Notice that Figure~\ref{fig:probplots1}(left) shows low confidence scores, while Figure~\ref{fig:probplots1}(right) shows sub-optimal performance for COVID-19 detection when using original training data only.

\begin{figure}[!htb]
\setlength{\tabcolsep}{.1em}
\centering
\begin{tabular}{cc}
\includegraphics[scale=.31]{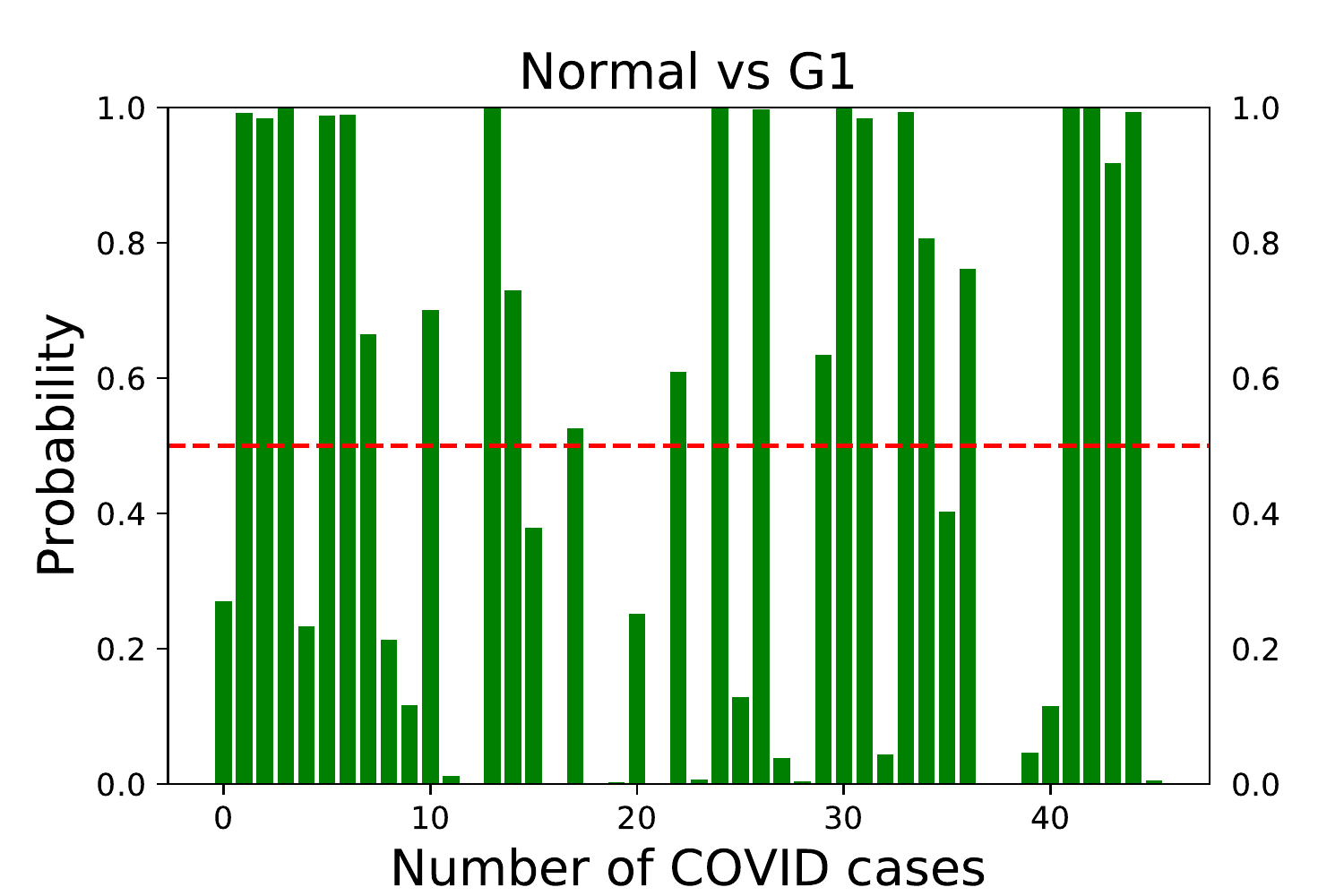} & \includegraphics[scale=.31]{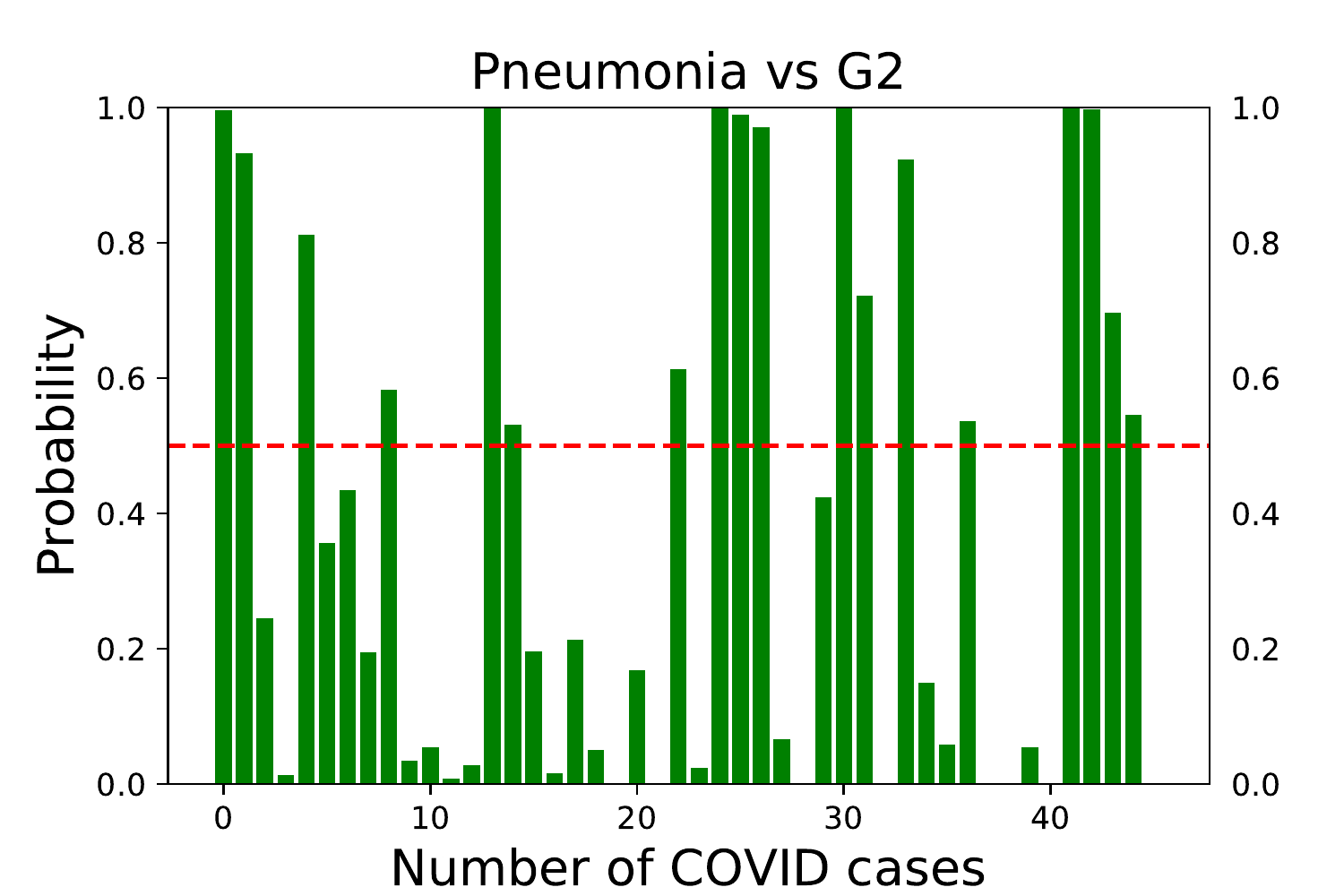}
\end{tabular}
\caption{Confidence scores for the VGG-16 model on unseen test set of COVID-19 for the two binary classification tasks of Normal vs. COVID-19 (left) and Pneumonia vs. COVID-19 (right) using synthetic data without the original examples. Notice that synthetic data increase the confidence scores.}
\label{fig:probplots2}
\end{figure}

Figure~\ref{fig:probplots2} shows that synthetic data can be used without the original examples. When using synthetic data as additional training set, we observe that not only the number of correctly detected instances of COVID-19 increases, but also the predictions tend to improve, as demonstrated by the high probability scores.

\begin{figure}[!htb]
\setlength{\tabcolsep}{.01em}
\centering
\begin{tabular}{cc}
\includegraphics[scale=.3]{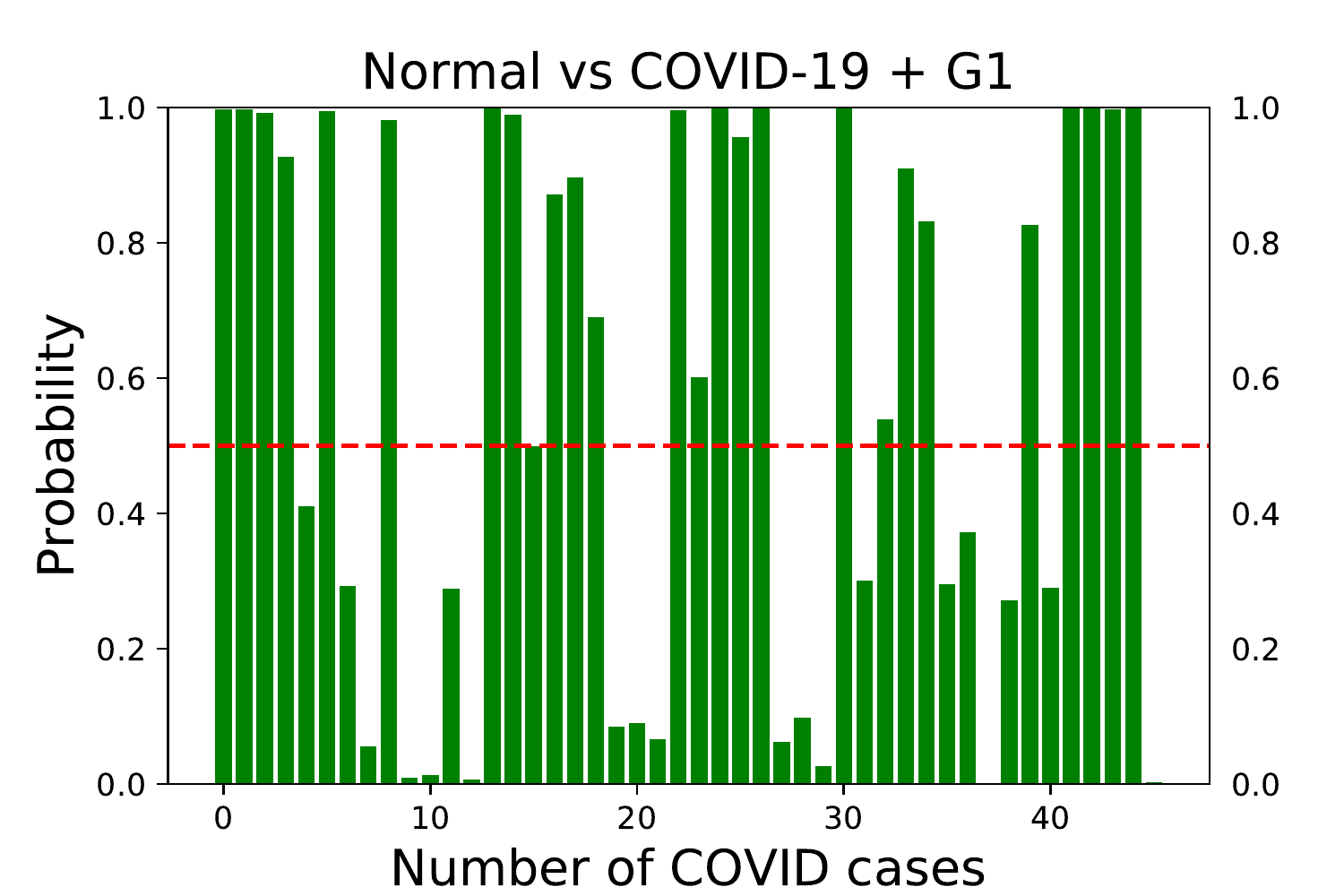} & \includegraphics[scale=.3]{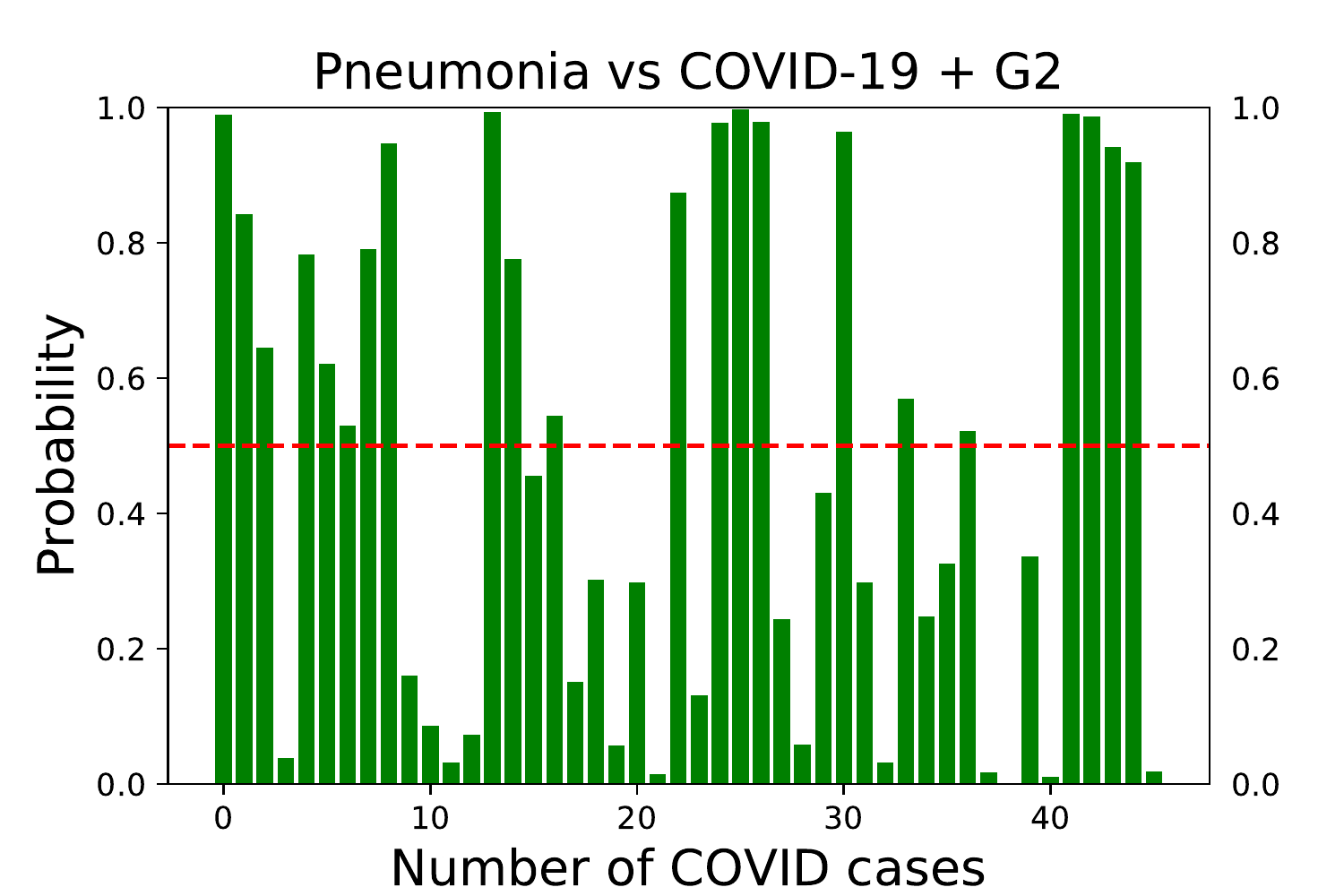}
\end{tabular}
\caption{Confidence scores for the VGG-16 model on unseen test set of COVID-19 for the two binary classification tasks of Normal vs. COVID-19 (left) and Pneumonia vs. COVID-19 (right) with synthetic data as additional training set. Left: adding 16,537 COVID-19 examples of $\mathcal{G}_{NC}$ to the original COVID-19 dataset. Right: adding 4,758 COVID-19 examples of $\mathcal{G}_{PC}$ to the original COVID-19 dataset.}
\label{fig:probplots3}
\end{figure}

\begin{figure}[!htb]
\setlength{\tabcolsep}{.01em}
\centering
\begin{tabular}{cc}
\includegraphics[scale=.3]{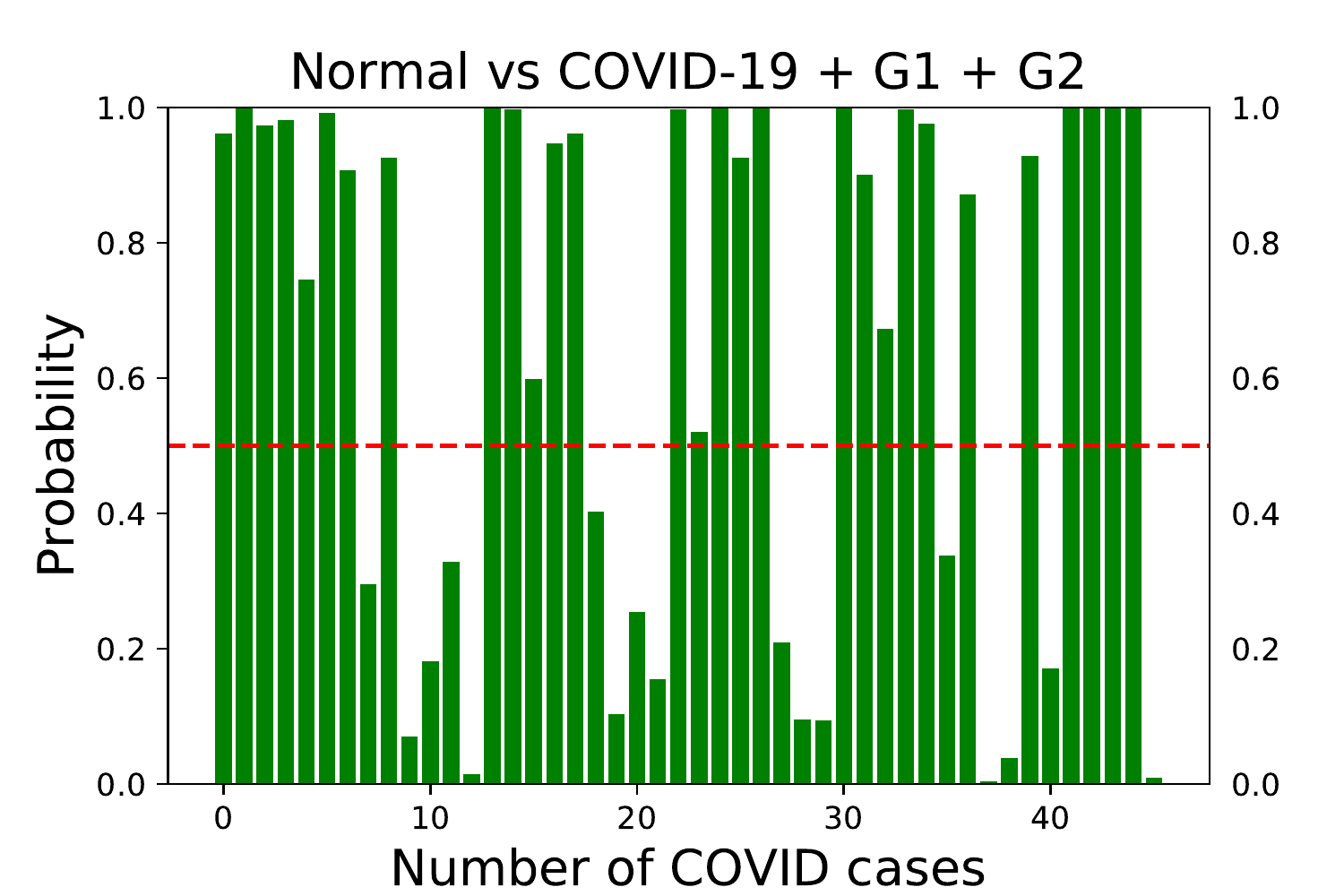} & \includegraphics[scale=.3]{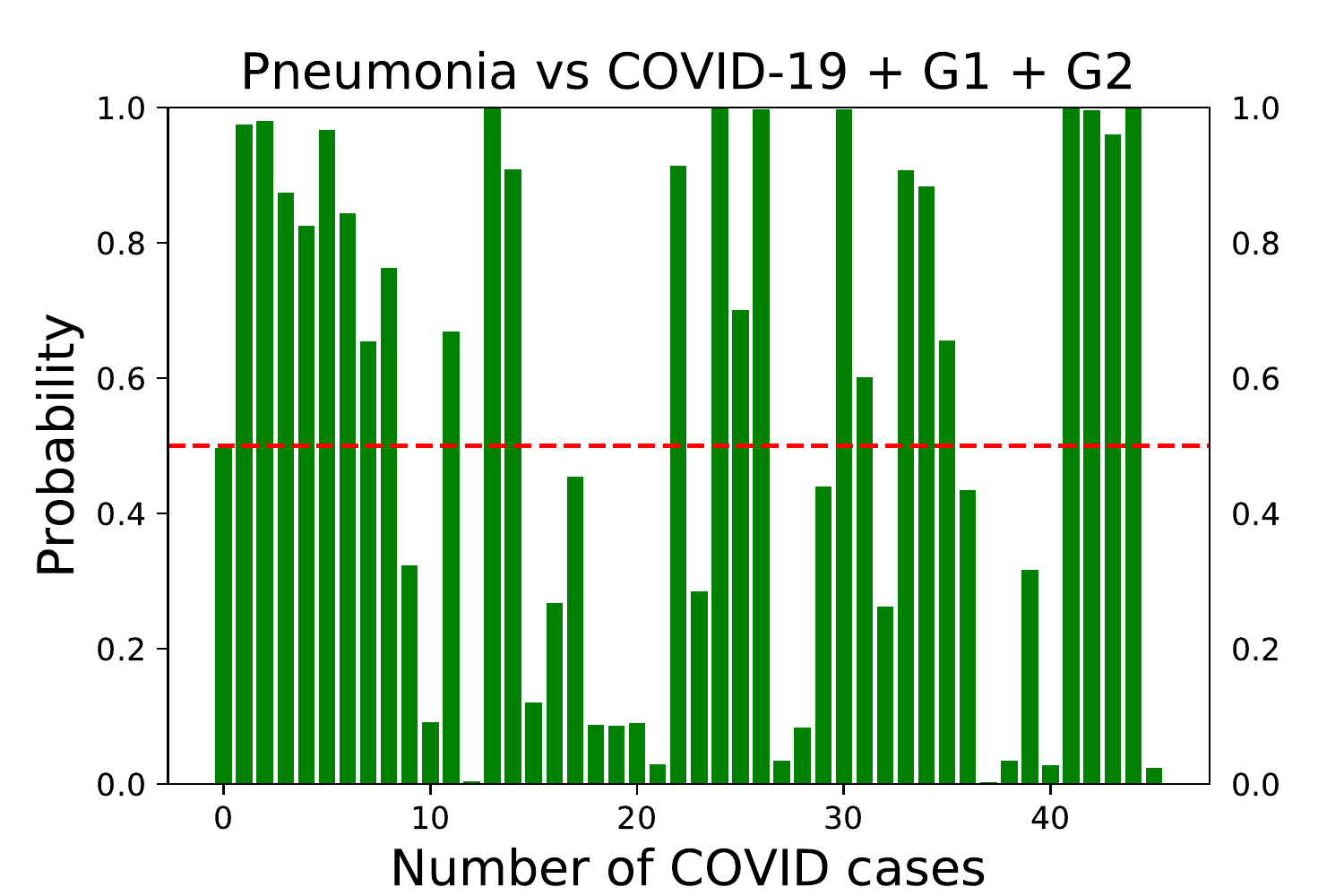}
\end{tabular}
\caption{Confidence scores for the VGG-16 model on unseen test set of COVID-19 for the two binary classification tasks of Normal vs. COVID-19 (left) and Pneumonia vs. COVID-19 (right) with synthetic data as additional training set. Both $\mathcal{G}_{NC}$ and $\mathcal{G}_{PC}$ are added to the original COVID-19 dataset.}
\label{fig:probplots4}
\end{figure}

Figures~\ref{fig:probplots3} and~\ref{fig:probplots4} show improved detection performance when the synthetic data are used as additional training set. A similar trend was observed with the ResNet-50 and DenseNet-102 models.

\subsection{Generating anonymized synthetic images with variation}
\begin{figure}[htb]
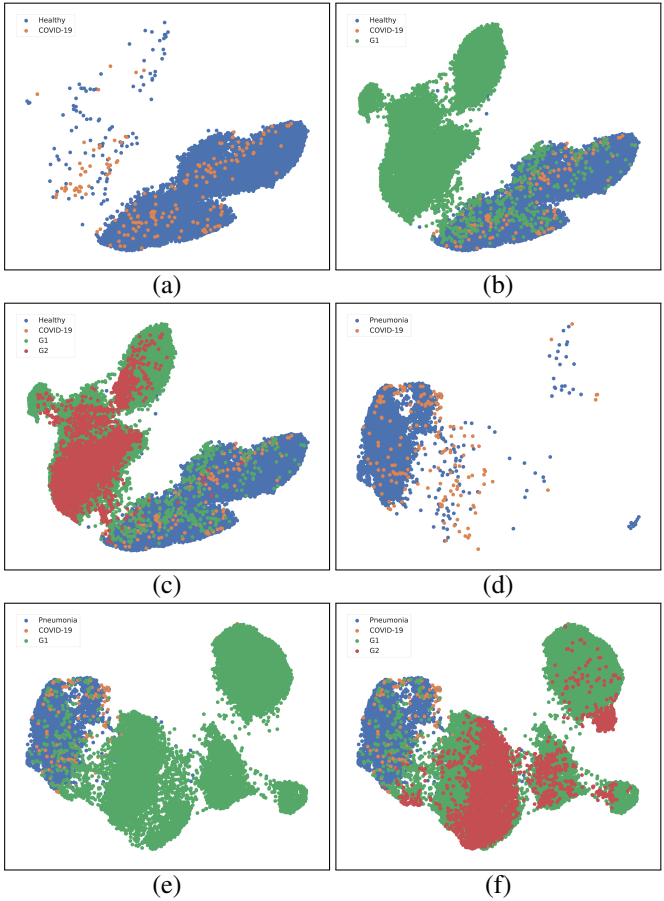

\setlength{\tabcolsep}{.1em}
\centering
\begin{tabular}{cc}
\fbox{\includegraphics[scale=.026]{Figure8a.pdf}} & \fbox{\includegraphics[scale=.026]{Figure8b.pdf}} \\
(a) & (b) \\
\fbox{\includegraphics[scale=.026]{Figure8c.pdf}} & \fbox{\includegraphics[scale=.026]{Figure8d.pdf}} \\
(c) & (d) \\
\fbox{\includegraphics[scale=.026]{Figure8e.pdf}} & \fbox{\includegraphics[scale=.026]{Figure8f.pdf}} \\
(e) & (f)
\end{tabular}
\caption{Two-dimensional UMAP embeddings: (a) Normal vs. COVID-19; (b) Normal vs. COVID-19 + $\mathcal{G}_{NC}$; (c) Normal vs. COVID-19 + $\mathcal{G}_{NC}$ + $\mathcal{G}_{PC}$; (d) Pneumonia vs. COVID-19; (e) Pneumonia vs COVID-19 + $\mathcal{G}_{NC}$; (f) Pneumonia vs. COVID-19 + $\mathcal{G}_{NC}$ + $\mathcal{G}_{PC}$. Here, G1 and G2 denote $\mathcal{G}_{NC}$ and $\mathcal{G}_{PC}$, respectively}
\label{fig:umaps}
\end{figure}

Data visualization based on dimension reduction plays an important role in data analysis and interpretation. The objective of dimension reduction is to map high-dimensional data into a low-dimensional space (usually 2D or 3D), while preserving the overall structure of the data as much as possible. A commonly used dimension reduction method is the Uniform Manifold Approximation and Projection (UMAP) algorithm, which is non-linear technique based on manifold learning and topological data analysis. UMAP is capable of preserving both local and most of the global structure of the data when an appropriate initialization of the embedding is used. The two-dimensional UMAP embeddings of the features are shown in Figure~\ref{fig:umaps} to visualize the difference between the original and synthetic data. Notice that the synthetic samples are in a different distribution in the feature space, enabling a decision boundary between the classes. The original examples in Figure~\ref{fig:umaps}(a) exhibit low inter-class variation and consist of outliers. In Figure~\ref{fig:umaps}(b), we can see that the synthetic examples of the $\mathcal{G}_{NC}$ dataset are in a different distribution in the feature space. While the UMAP embeddings may not be interpreted as a justification that the synthetic examples actually consist of COVID-19 symptoms from a clinical perspective, it is, however, important to note that the distribution of the synthetic images is significantly different than that of normal images; thereby enabling a proper decision boundary. A similar trend can be observed in Figures~\ref{fig:umaps}(d), (e) and (f). The overlapping features for Pneumonia vs. COVID-19 can be explained by the fact that the findings of X-ray imaging in COVID-19 are not specific, and tend to overlap with other infections such as Pneumonia in this case.

\subsection{Discussion}
Since the generative and classification models are trained to learn representations in the training data distribution, it is likely that a bias might occur toward that data. In light of the class imbalance problem, the generator is trained by under-sampling the majority class. This under-sampling process often leaves a relatively small number of data points (180 samples for each domain) to learn from. While a boost in performance is achieved when using the synthetic datasets, it is not conclusive enough to confirm whether our approach can be generalized across other COVID-19 datasets due largely to the lack of such benchmarks. While the improvements we have achieved using our proposed framework are encouraging, it is important to mention that a key objective of this work is not to claim state-of-the-art results, but rather to release an open source dataset to the research community in an effort to further improve COVID-19 detection.

\section{Conclusion}
In this paper, we presented an unsupervised domain adaptation approach by leveraging class conditioning and adversarial training to build an open database of synthetic COVID-19 chest X-ray images of high fidelity. This publicly available database comprises 21,295 synthetic images of chest X-rays for COVID-19 positive cases. The insights generated from applying recent deep learning approaches on this database can be used for preventive actions against the global COVID-19 pandemic, in the hope of containing the virus. We also demonstrated how the data generation procedure can serve as an anonymization tool by achieving comparable detection performance when trained only on synthetic data versus real data in an effort to alleviate data privacy concerns. Our experiments reveal that synthetic data can significantly improve the COVID-19 detection performance results, that as the amount of synthetic data is increased, sensitivity improves considerably and the number of false negatives decreases. We believe that the performance can be further improved by applying more application-specific preprocessing and exhaustive hyper-parameter tuning, as well as by leveraging ensemble methods, which we leave for future work.

\section{Acknowledgements}
This work was supported in part by and Natural Sciences and Engineering Research Council of Canada (NSERC) Discovery Grant Number N00929. This research was enabled in part by advanced computing resources provided by Compute Canada.


\bibliographystyle{ieeetr}
\bibliography{references}
\end{document}